\documentclass[twoside,english]{paper}
\usepackage[T1]{fontenc}
\usepackage[latin9]{inputenc}
\usepackage{geometry}
\usepackage{rotating}
\usepackage{textcomp}
\usepackage{amstext}
\usepackage{graphicx}

\makeatletter

\providecommand{\tabularnewline}{\\}

\newcommand{\lyxaddress}[1]{
\par {\raggedright #1
\vspace{1.4em}
\noindent\par}
}

\usepackage[numbers, sort&compress]{natbib}

\makeatother

\usepackage{babel}
\begin{document}

\title{Resonant multi-photon IR dissociation spectroscopy of a trapped and
sympathetically cooled biomolecular ion species}

\author{Ch. Wellers%
\thanks{First authorship shared. This is an originally submitted version. Revised version was accepted to the Physical Chemistry Chemical Physics.%
} , A. Borodin$^{*}$, S. Vasilyev, D. Offenberg and S. Schiller}

\maketitle

\lyxaddress{Institut für Experimentalphysik, Heinrich-Heine-Universität Düsseldorf,
40225 Düsseldorf, Germany}
\begin{abstract}
In this work we demonstrate vibrational spectroscopy of polyatomic
ions that are trapped and sympathetically cooled by laser-cooled atomic
ions. We use the protonated dipeptide tryptophane-alanine (HTyrAla$^{+}$)
as a model system, cooled by Barium ions to less than 800\,mK secular
temperature. The spectroscopy is performed on the fundamental vibrational
transition of a local vibrational mode at 2.74\,\textmu{}m using
a continuous-wave optical parametric oscillator (OPO). Resonant multi-photon
IR dissociation spectroscopy (without the use of a UV laser) generates
charged molecular fragments, which are sympathetically cooled and
trapped, and subsequently released from the trap and counted. We measured
the cross section for R-IRMPD under conditions of low intensity, and
found it to be approximately two orders smaller than the vibrational
excitation cross section. 

The observed rotational bandwidth of the vibrational transition is
larger than the one expected from the combined effects of 300 K black-body
temperature, conformer-dependent line shifts, and intermolecular vibrational
relaxation broadening (J. Stearns et al., J. Chem. Phys., \textbf{127,
}154322-7 (2007)). This indicates that as the internal energy of the
molecule grows, an increase of the rotational temperature of the molecular
ions well above room temperature (up to on the order of 1000\,K),
and/or an appreciable shift of the vibrational transition frequency
(approx. 6-8\,cm$^{-1}$) occurs.
\end{abstract}

\section{Introduction}

In the field of cold, trapped molecules, one topic under current study
is the development and refinement of suitable spectroscopic methods.
Concerning more specifically translationally cold molecular ions,
often produced via sympathetic cooling by laser-cooled ions, a few
studies have been performed recently, on diatomic molecular ions.
Ro-vibrational spectroscopy is one example, applied in particular
to HD$^{+}$ and MgH$^{+}$. Fundamental ($\Delta v=1$) \cite{Schneider2010,Staanum,Vogelius,Nielsen},
first ($\Delta v=2$) \cite{Schneider2010}, and third overtone ($\Delta v=4$)
\cite{Roth2006,Koelemeij2007} vibrational spectroscopy, with resolution
of the rotational structure, has been described. In the simplest case,
the spectroscopy consisted in a simple variant of resonance-enhanced
multiphoton dissociation (1+1' REMPD), where a vibrational transition
is driven by one photon and a second, UV photon dissociates the molecule
by exciting it from the vibrationally excited level to a non-binding
electronically excited state. The reduction of the number of intact
molecular ions is then detected in an appropriate way. More recently,
in our group we have also employed 1+1'+1'' REMPD, where the vibrationally
excited molecule was further vibrationally excited before being dissociated
by a UV photon \cite{Bressel}. Finally, pure rotational spectroscopy
could also be demonstrated by us, again via 1+1'+1'' REMPD, on HD$^{+}$.
It is of interest to explore if and how spectroscopy can be extended
to sympathetically cooled molecular ions other than diatomic ones.

This work is devoted to a first methodological study of vibrational
spectroscopy of a sympathetically cooled polyatomic molecular ion.
Vibrational spectroscopy of complex molecular ions is a very important
method for gaining information on structures of molecules, including
biomolecules (see, e.g.\cite{Correia2008} and references therein),
and for studying the energetics of fragmentation \cite{Dienes1996}.
Apart from implementing vibrational spectroscopy, we investigate the
question whether the rotational degrees of freedom of the molecules
couple, via the ion-ion collisions between the molecules and other
ions in the trap, to the translational ones, which are sympathetically
cooled to less than 800\,mK. The molecular ion species used, protonated
tyrosine-alanine (HTyrAla$^{+}$), is a test case appropriate for
the present purposes since a previous study \cite{Stearns2007} has
shown that it exhibits a spectrally narrow vibrational transition.
The vibrational linewidth included the width of the rotational band,
whose details were unresolved, due to experimental reasons, but also
because of the small rotational constant exhibited by this relatively
large molecule. 

We found in our study that a different spectroscopic technique as
compared to REMPD is applicable on this test ion, namely resonant
multi-photon IR dissociation (R-IRMPD). Here, only photons of energy
resonant with a vibrational transition of a specific vibrational mode
are used. Upon absorption of the photon, the vibrational energy is
distributed on a time scale of less than a nanosecond \cite{Gregoire2007}
to other internal degrees of freedom, by intramolecular vibrational
relaxation (IVR). After that, another photon of the same energy can
be absorbed on the same transition \cite{Black1979,Polfer2009}. This
is not possible with sympathetically cooled diatomic ions, since there
are no other internal degrees of freedom, nor sufficient collisional
vibrational relaxation because UHV conditions are maintained. The
photon absorption process is repeated, allowing the molecule to collect
internal energy until it will be large enough to break certain molecular
bonds, leading to generation of charged fragments, and a reduction
in the number of parent molecules. This technique is used in spectroscopy
of trapped complex molecular ions, but has so far been applied only
to uncooled molecules, to our knowledge. It is used in conjunction
with large-scale, pulsed infrared laser systems, e.g. free-electron
lasers, which allow a very wide spectral coverage \cite{Correia2008,Duarte1995}
and therefore a rather complete study of the vibrational spectrum. 

The present laboratory experiment is an example that such studies
are becoming possible with laboratory laser sources only, even continuous-wave
ones. Indeed the IR spectral coverage with cw sources such as OPOs,
quantum cascade lasers and difference frequency generators \cite{Vasilyev2010}
is rapidly increasing.

\section{Experimental apparatus\label{sec:Experimental-apparatus}}

\begin{figure}
\begin{centering}
\includegraphics[width=12cm]{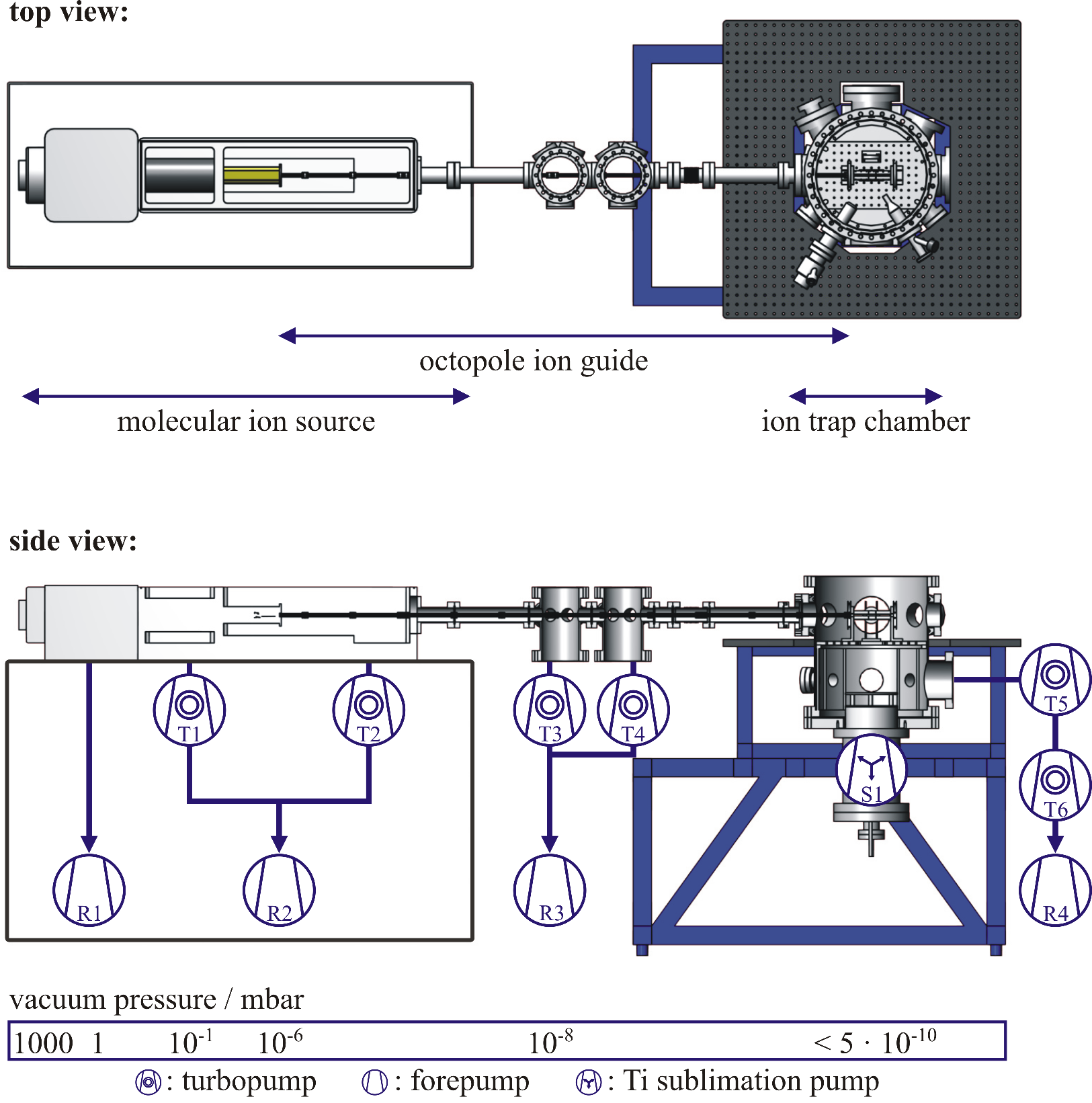}
\par\end{centering}

\caption{Overview of the vacuum setup. Top and side view of the vacuum setup
consisting of the electrospray ionization (ESI) molecular ion source,
the octopole ion guide, and the ion trap vacuum chamber in which the
molecular ions are trapped, cooled and manipulated in a linear quadrupole
ion trap. The vacuum pressure in the differential vacuum setup decreases
from atmospheric pressure at the molecular ion source inlet (where
a solution is fed into the ion source) to ultra-high vacuum values
in the ion trap chamber \cite{Ostendorf2004,Offenberg2009}.}

\label{Flo:vaccum setup}
\end{figure}

In this work, a previously developed, general-purpose apparatus for
sympathetic cooling of a large variety of complex molecular ions is
employed \cite{Ostendorf2006}. Figure \ref{Flo:vaccum setup} shows
an overview of the apparatus.
 The molecular ions travel from left, the ion source, to the right,
the ion trap. Mass-to-charge selected molecular ions are trapped in
the ion trap and are sympathetically cooled via Coulomb interaction
with a laser-cooled atomic ion ensemble species. Depending on their
mass and charge, polyatomic ions can be cooled to translational temperatures
well below 1\,K \cite{Offenberg2008}. The long storage time of few
tens of minutes combined with the well-defined and nearly collisionless
environment in the ion trap, provides very particular conditions for
spectroscopy.

As the electrospray ionization takes place at atmospheric pressure,
while the achievement of sub-Kelvin temperature using laser cooling
requires ultra-high vacuum conditions, high demands are made on the
vacuum system. Both conditions are fulfilled using a differentially
pumped vacuum setup that provides a pressure gradient of more than
13 orders of magnitude, sustained by in total 11 vacuum pumps. 

\begin{figure}
\begin{centering}
\includegraphics[width=9cm]{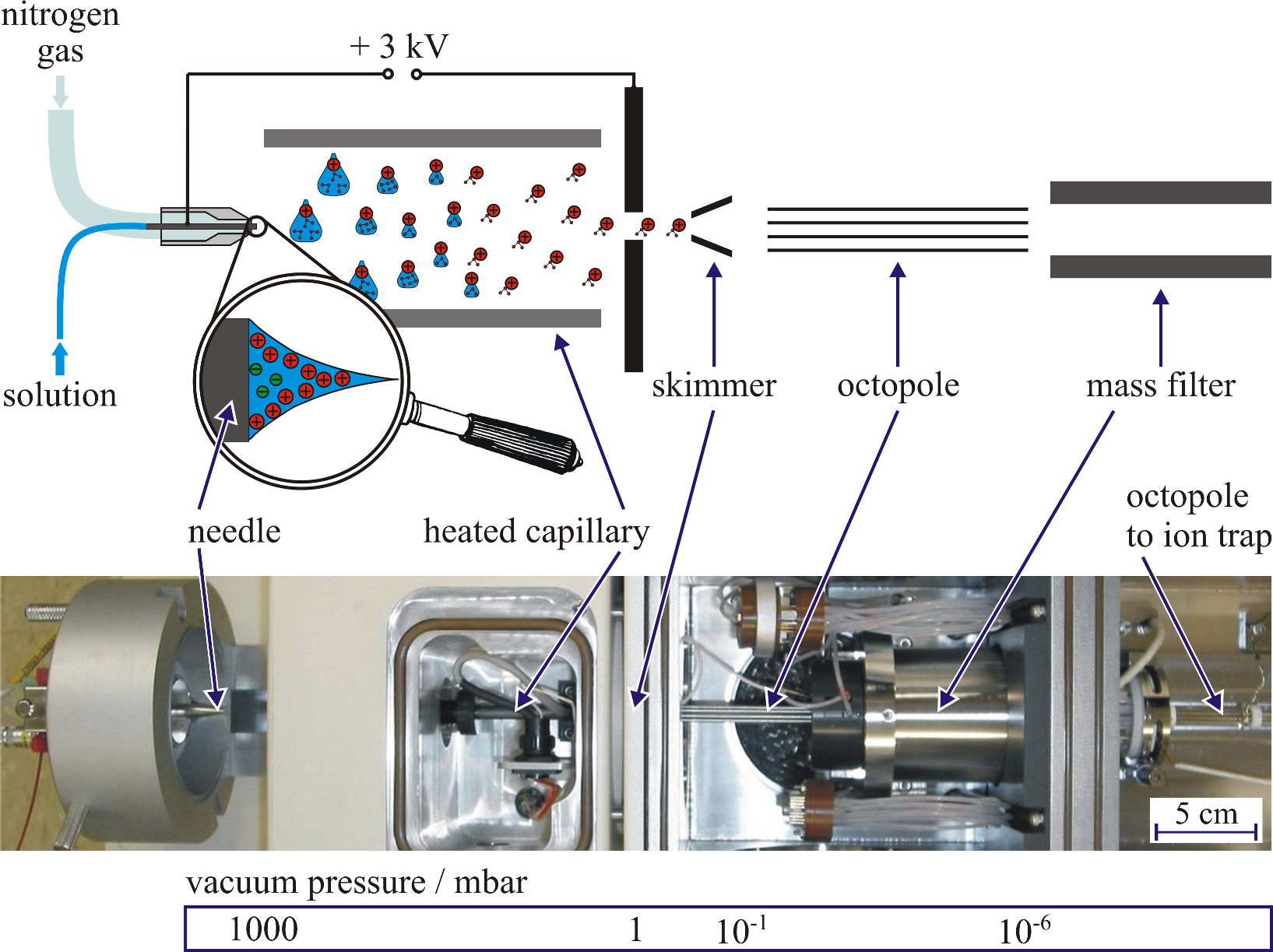}
\par\end{centering}

\caption{Principle and setup of the molecular ion source. The upper schematic
diagram (not to scale) illustrates the principle of electrospray ionization
as described in the text. The lower photo shows the uncovered molecular
ion source, with the needle through which the solution with the molecular
ions is sprayed, the heated capillary in which the solvent evaporates,
the skimmer (covered by housing components) which serves as aperture,
a short octopole used as ion guide, the quadrupole mass filter for
selection of the ion species to be transferred to the ion trap, and
the beginning of the 2 m long octopole which transfers the molecular
ions to the ion trap. The pressures given below the photograph are
approximate values for the atmospheric pressure region (1000\,mbar),
the skimmer region (1\,mbar), the octopole region ($10^{-1}$ \,mbar),
and the following high vacuum region ($10^{-6}$\,mbar).}

\label{Flo:ESI}
\end{figure}

The molecular ion source is a modified commercial mass spectrometer
Finnigan SSQ\,700, based on electrospray ionization (ESI). A large
variety of neutral molecules can be protonated (i.e. one or more proton(s)
attach(es) to the neutral molecule by polarization interaction) in
the ESI molecular ion source, including molecules with masses exceeding
10,000 atomic mass units (amu). As (Figure \ref{Flo:ESI}) 
 shows, the molecular ions are produced from a solution, which is
sprayed as droplets into a vacuum chamber, where they evaporate the
solvent until individual protonated molecular ions remain. 

\begin{figure}
\begin{centering}
\includegraphics[width=15cm]{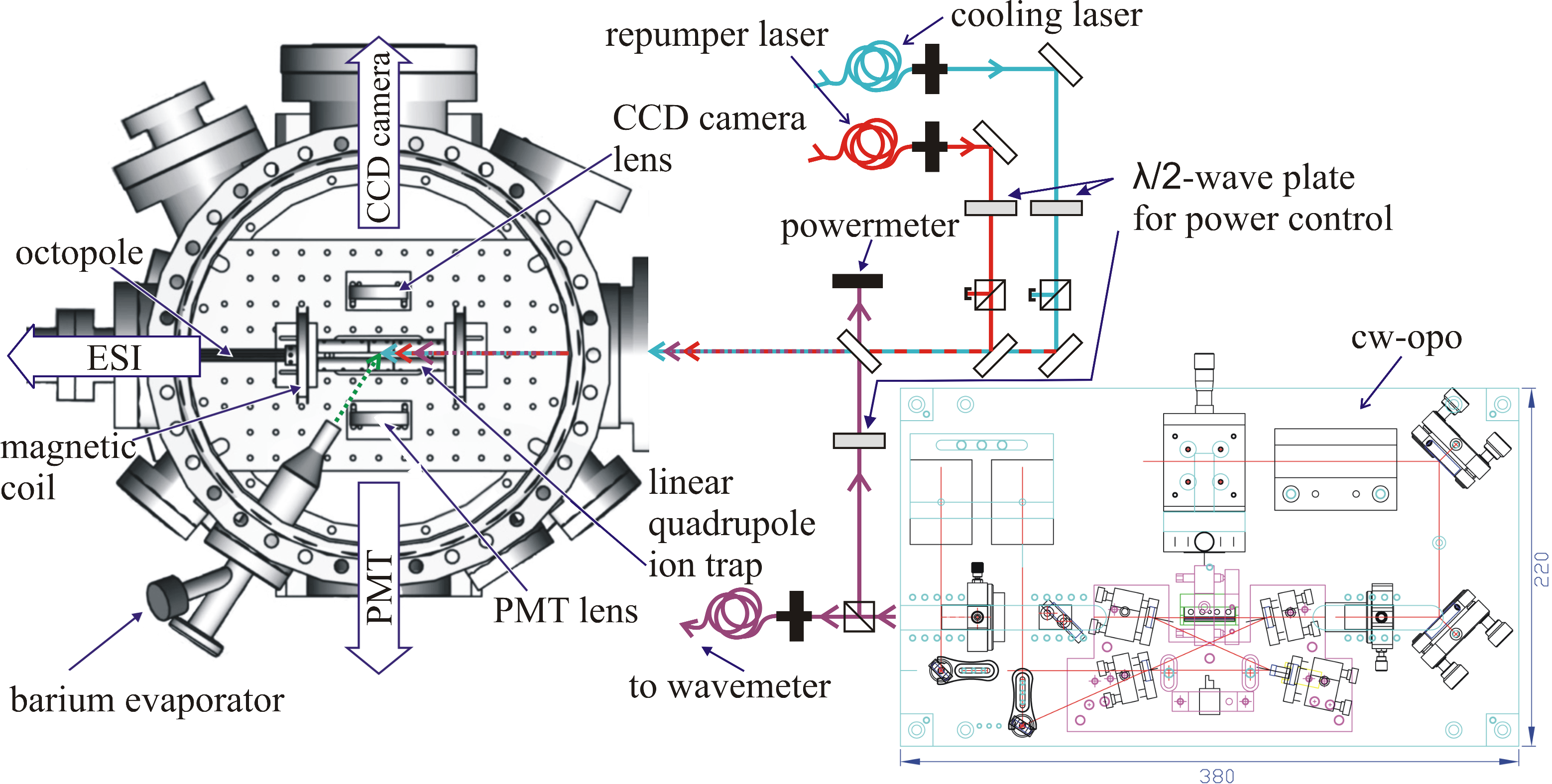}
\par\end{centering}

\caption{Overview of the ion trap vacuum chamber. Molecular ions from the ESI
ion source are transferred to the linear quadrupole ion trap via the
octopole ion guide. The barium ions are produced with an ultra-high-vacuum
suitable evaporator directed to the trap center (dashed green arrow).
Cooling (493\,nm), repumping (650\,nm) and vibrational spectroscopy
laser light (2.7\,\textmu{}m) (blue, red and violet arrows, respectively)
enter the vacuum chamber through an axial sapphire viewport. The fluorescence
light emitted by the laser-cooled $^{138}$Ba$^{+}$ ions is collimated
by radially arranged lenses to direct it through glass viewports to
a CCD camera and a photomultiplier (PMT), both located outside the
vacuum chamber \cite{Ostendorf2004}. The dimensions of the OPO base
plate are in mm. }

\label{Flo:overview trap}
\end{figure}

In the present study, a solution of 70\,pmol/l dipeptide tyrosine-alanine
in a solution of methanol + water (1:1 relative concentration) + 0.1\%
formic acid is prepared, in which protonated tyrosine-alanine forms
(HTyrAla$^{+}$, mass 251\,amu). The protons are supplied by the
formic acid. The solution is filled into a syringe, from which it
is sprayed at atmospheric pressure (1000\,mbar) through a thin metal
needle to which a positive high voltage is applied. The emerging solution
is shaped by a streaming nitrogen gas into a so-called Taylor cone.
Solution droplets break away from the Taylor cone and are accelerated
towards a heated capillary (200\,\textdegree{}C). The solvent evaporates
and the droplets shrink until a critical size, known as Rayleigh stability
limit, is reached. The droplets then break apart into smaller ones
until bare molecular ions remain. The stream of molecular ions is
transferred through the skimmer region at 1 mbar to the quadrupole
mass filter with a short octopole ion guide. This mass filter is placed
in the high-vacuum section of the apparatus, at about 10$^{\text{-6}}$\,mbar
pressure. The quadrupole mass filter parameters are controlled by
computer, allowing purifying the ion stream and delivering a species
with a particular mass-to-charge ratio to the following apparatus
section. After the mass filter, the molecular ions are coupled into
a 2 m long radio frequency octopole ion guide which transfers the
ions to the linear quadrupole ion trap placed in a cylindrical vacuum
chamber at < 10$^{-9}$\,mbar, shown in Figure \ref{Flo:overview trap}.
 Details about the octopole guide can be found in \cite{Borawski2004,Ostendorf2004,Zhang2008}. 

The RF trap consists of four cylindrical stainless steel rods with
a diameter of 10\,mm, which are divided into the middle electrodes
of 20\,mm length and the two endcap electrodes of 30\,mm length.
The smallest distance of the electrode surface to the trap axis is
$r_{0}=4.36\,\mathrm{mm}$. With this geometry, the axial electric
potential of the trap \cite{Hornekaer2000,Drewsen2000}, 
\[
\Phi_{z}=\kappa U_{EC}z^{2},
\]
has the coefficient $\kappa=1500\,\mathrm{m}^{-2}$. The RF voltages
for radial ion confinement are applied to the middle and the endcap
electrodes with an RF frequency of $\Omega=2\pi\cdot2.5\,\mathrm{MHz}$
and a maximum amplitude of $U_{RF,max}=500\,\mathrm{V}$. DC voltages
of $-20\,\mathrm{to}\,+20\,\mathrm{V}$ can be individually applied
to the four middle electrodes to compensate external electric fields
for exact symmetrization of the quadrupole field. Typical operation
parameters are $200\,\mathrm{V}<U_{RF}<500\,\mathrm{V}$ and $5\,\mathrm{V}<U_{EC}<7\,\mathrm{V}$.

\begin{figure}
\begin{centering}
\includegraphics[width=9cm]{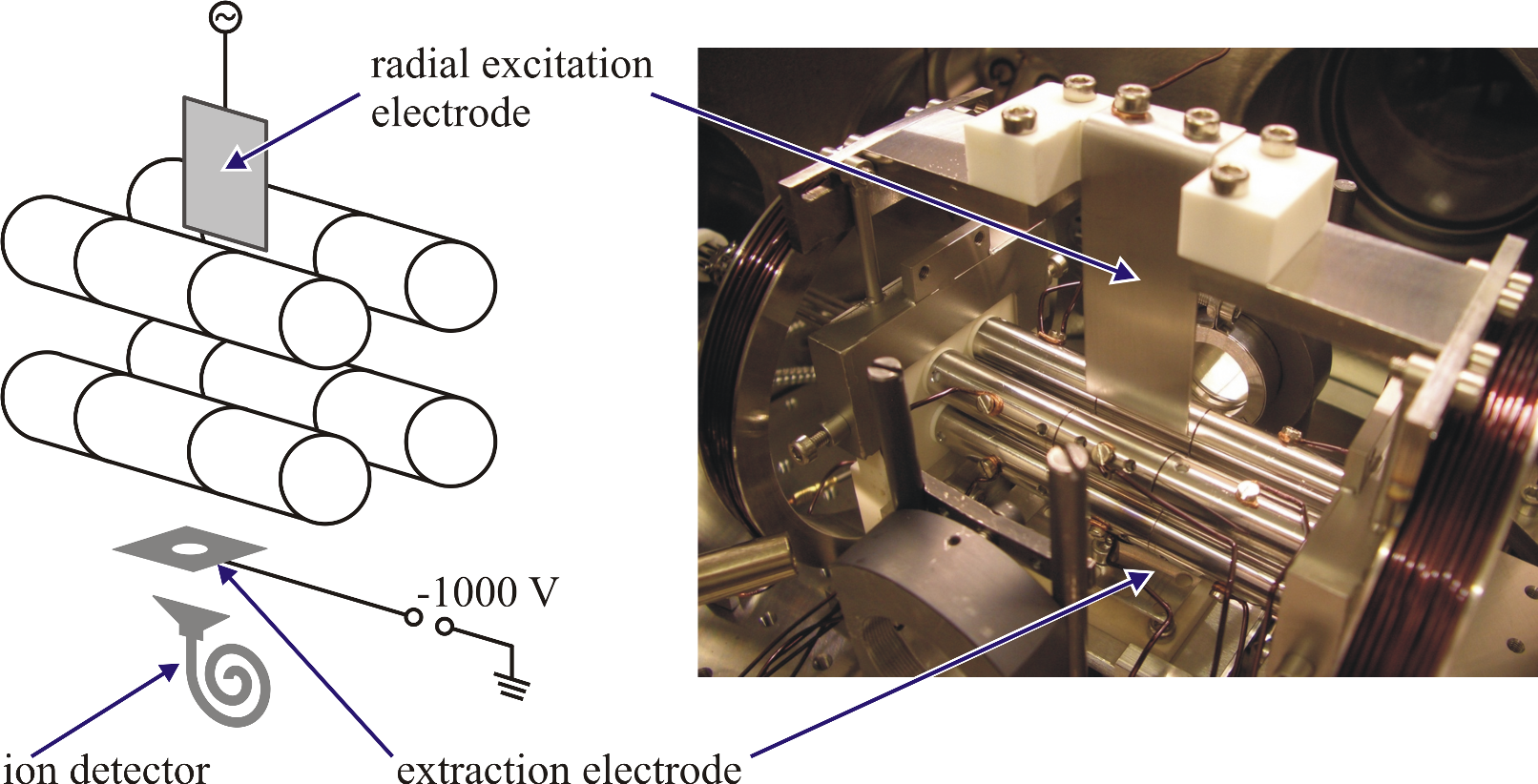}
\par\end{centering}

\caption{Trap setup with cleaning and detection components. With an AC voltage
scan applied to the radial excitation electrode, molecules with a
specified mass-to-charge ratio can be removed from the trap. The identification
of the stored ion species is via their mass-to-charge ratio, and is
obtained via a controlled decrease of the ion trap RF amplitude and
a simultaneous detection of the escaping ions. Detection is achieved
by directing the ions by an extraction electrode to an ion detector
mounted below the trap.}

\label{Flo:Trap setup}
\end{figure}

The magnetic field that is necessary for laser cooling of $^{\text{138}}$Ba$^{\text{+}}$
ions is produced by two coils centered around the trap axis and mounted
beyond the endcap electrodes (see Figure \ref{Flo:Trap setup}). The
coils produce a magnetic field in the trap center which is parallel
to the trap axis and has a strength of approximately 5\,G.

To count the trapped ions, they a released from the trap and led to
a channel electron multiplier (CEM) mounted below the trap, serving
as an ion detector (Figure \ref{Flo:Trap setup}). 
 When the trap RF voltage $U_{RF}$ is decreased, ions of mass $m$
and charge $q$ are not stably trapped any more once the value reaches
\[
U_{RF}^{ex}=ar_{0}^{2}\text{\ensuremath{\Omega\sqrt{2\kappa\mbox{\ensuremath{\mathit{\mathit{U}}_{\mathit{EC}}\frac{\mathit{m}}{\mathit{q}}}}}}}
\]
Here $a$ is a numerical factor describing the influence of the electric
field of the extraction electrode. This simple expression holds in
the limit of a single trapped ion and zero kinetic energy.

The escaping ions will be accelerated by an extraction electrode at
a DC voltage of about -1000\,V and counted. The number of detected
ions versus the RF voltage represents a mass spectrum of the trapped
ion ensembles ($q=e$ is assumed for all ions). In the spectra shown
below, the largest detectable mass (850\,amu) corresponds to the
rf amplitude used for normal trapping operation, $U_{RF}^{ex}=U_{RF}=300\,$V.
By comparing the ion number detected by the counter with the number
of ions initially contained in the Coulomb crystals (this number is
obtained by Molecular Dynamic (MD) simulations on the basis of CCD
images), we found that the ion detection efficiency is between 10
and 20\,\%.

For the laser cooling of $^{138}$Ba$^{+}$ ions, a cooling laser
at 493\,nm (generated by sum-frequency mixing of a 920\,nm diode
laser wave and 1064\,nm Nd:YAG laser wave) and a repumper diode laser
at 650\,nm are required. Both cooling and repumping laser frequencies
are locked to transfer cavities, which, in turn, are locked to the
Nd:YAG laser, itself stabilized to a ro-vibrational transition of
molecular iodine via Doppler-free saturation spectroscopy.

\section{Preparation, cooling and analysis of ion ensembles}

\begin{figure}
\begin{centering}
\includegraphics[width=15cm]{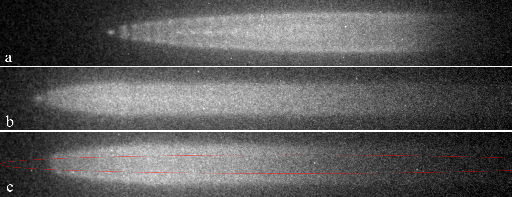}
\par\end{centering}

\caption{Experimental CCD images of ion crystals.\textbf{ a)} Barium ion crystal.
\textbf{b)} Same crystal, surrounded by invisible HTyrAla$^{+}$ions.
\textbf{c)} Crystal after partial photodissociation of HTyrAla$^{+}$ions
by R-IRMPD. Molecular fragments lighter than Ba$^{+}$ ions are trapped
close to the trap axis, leading to a reduction of the fluorescence
of the Ba$^{+}$ ions in that region, which is marked in red.}

\label{Flo:crystal}
\end{figure}

\begin{figure}
\begin{centering}
\includegraphics[width=9cm]{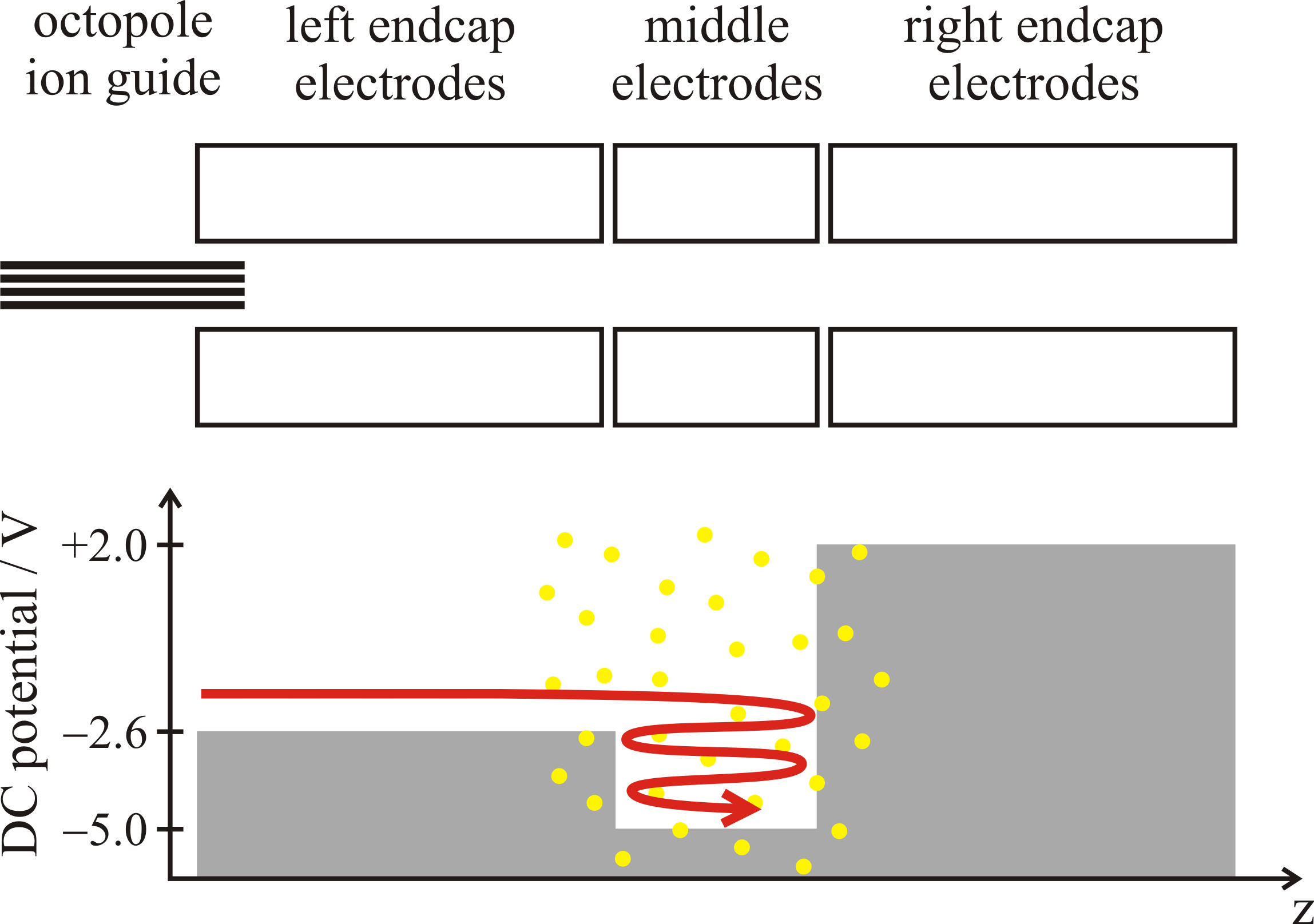}
\par\end{centering}

\caption{Schematic principle of loading of the ion trap with molecular ions
coming from the ion source. The DC potential of the left endcap electrodes
is temporarily lowered, allowing entrance of the ions. Their kinetic
energy (red arrow) is reduced by collisions with helium buffer gas
(yellow dots), leading to trapping.}

\label{Flo:potential trap}
\end{figure}

First, a barium ion crystal is prepared in the trap (\cite{BRoth2005},
Figure \ref{Flo:crystal} a)
). The source of barium ions is an oven, which typically also produces
Barium oxide ions. Both species are trapped in the ion trap. Laser-cooling
cooles the atomic ions directly, some BaO$^{+}$ ions are sympathetically
cooled by the atomic ions. In order to load molecular ions delivered
via the octopole ion guide, the DC potential of the entry endcap is
decreased and the ion guide is activated for several seconds. In addition,
room temperature helium buffer gas (at $4\ldots6\cdot10^{-6}$\,mbar)
is injected into the trap using a piezoelectric valve. This acts as
a buffer gas, in which collisions with the molecular ions reduce their
kinetic energy sufficiently for capture in the trap (Figure \ref{Flo:potential trap}
). The ions in the crystal also collide with the He atoms, causing
its melting. The continued action of the cooling laser leads to a
steady state corresponding to a cold gas. After restoring the entry
electrodes' potential the valve is closed and the pressure of the
buffer gas decreases within several seconds due to pumping. Then the
Barium cloud crystallizes again and sympathetically cools the trapped
molecular ions (Figure \ref{Flo:crystal} b)). For species-selective
cleaning (removal of unintended trapped BaO$^{+}$ and CO$_{2}^{+}$
ions) we apply a suitably strong AC electric field to a radial excitation
electrode (see Figure \ref{Flo:Trap setup}). It induces a radial
secular motion of sufficiently large amplitude so that the resonant
ions are ejected from the trap. The frequencies of the exciation are
made to cover the secular resonance frequencies of undesired ions.
The cleaning is used in particular to to remove CO$_{2}^{+}$ and
BaO$^{+}$.

\begin{figure}
\begin{centering}
\includegraphics[width=7.5cm]{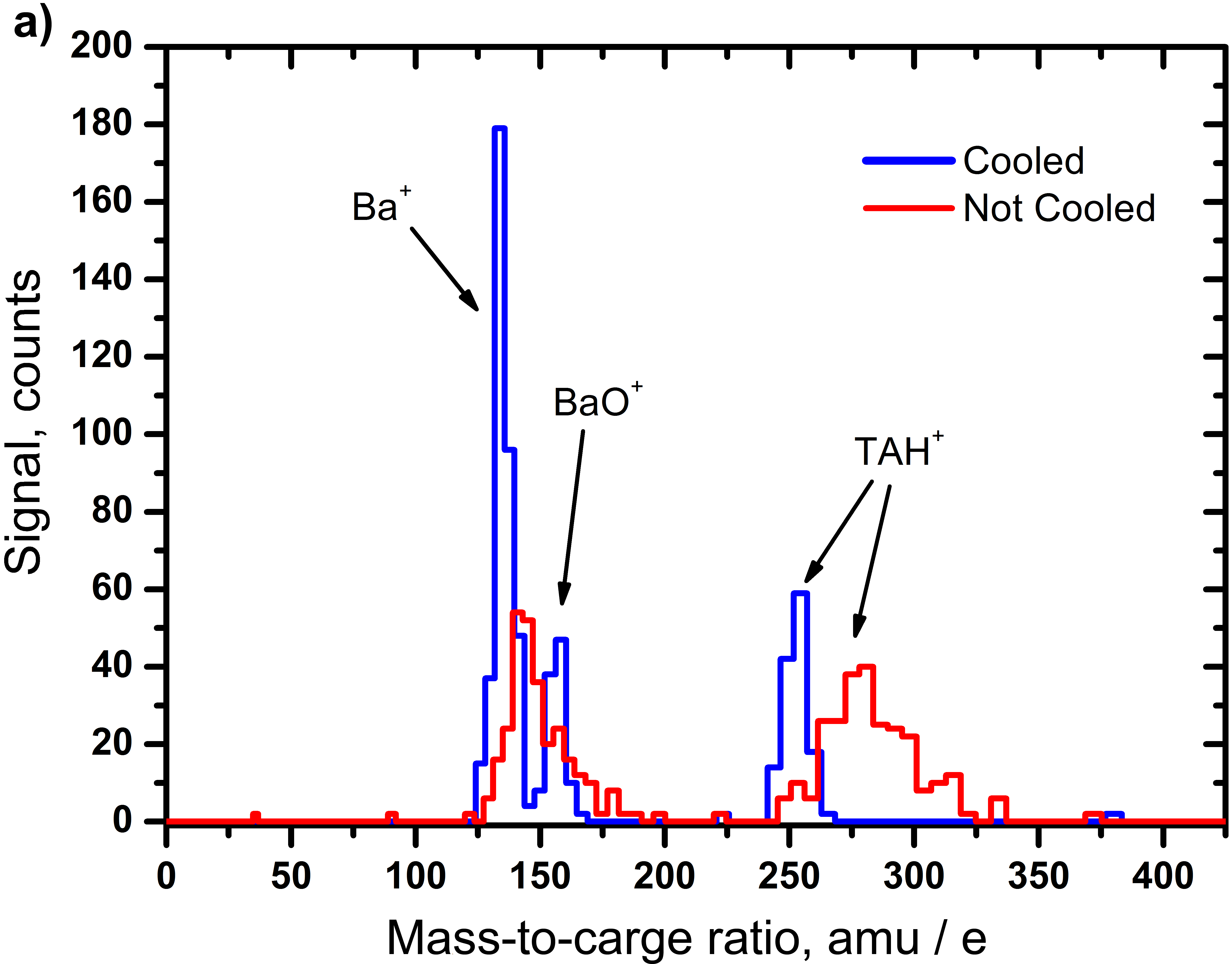}\includegraphics[width=7.5cm]{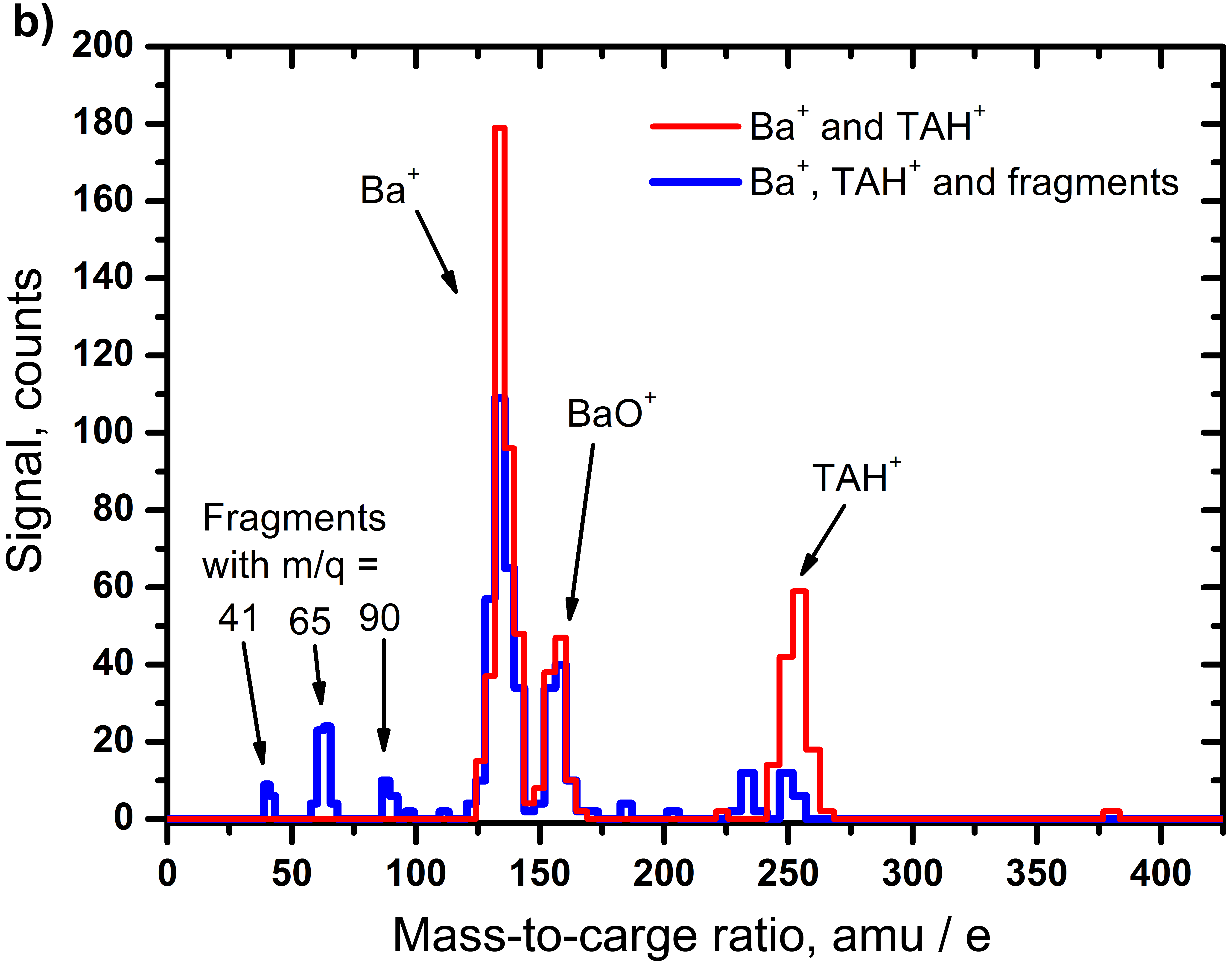}
\par\end{centering}

\caption{\textbf{a)} Mass spectrum of extracted ions, when the trapped ensemble
was laser-cooled (blue) and buffer-gas cooled (300\,K, red), respectively.
The wider peaks in the 300\,K case are due to the correspondingly
larger kinetic energy distribution of the ions inside the trap. Note
also that in the 300\,K case, ions are extracted from the trap already
at higher RF amplitudes as compared to cold ions. \textbf{b)} Mass
spectrum of a cold ion ensemble without R-IRMPD (red) and with R-IRMPD
(blue). }

\label{Flo:Hot vs. cold a}\label{Flo:extraction b}
\end{figure}

Two ion extraction spectra obtained as described in Sec.\ref{sec:Experimental-apparatus}
are compared in Figure \ref{Flo:Hot vs. cold a} a)
 to show the significant influence of laser cooling and sympathetic
cooling. The ensembles contained Ba$^{+}$, BaO$^{+}$ and HTyrAla$^{+}$
and had kinetic energies corresponding to room temperature (red),
achieved by collisions with He buffer gas (lasers were off), and below
1 K (blue), achieved by laser cooling. In the 300\,K case, the peaks
in the extraction spectrum are broadened, indicating a larger distribution
in the energy of the trapped ions. For the cooled ensemble, not only
the peak of the laser-cooled Ba$^{+}$ ions is narrower, allowing
to resolve the peak of the sympathetically cooled BaO$^{+}$ions,
but also the one of HTyrAla$^{+}$. This latter peak narrowing is
a qualitative proof of the sympathetic cooling of complex molecular
ions. 

The action of R-IRMPD on the ion ensemble is seen in Figure \ref{Flo:crystal}
c), where a Ba$^{+}$/HTyrAla$^{+}$ ion crystal was exposed comparatively
long to continuous-wave light at 2.74\,\textmu{}m. Figure \ref{Flo:Hot vs. cold a}
b) shows the ion extraction spectra, which gives detailed information
about the contents of the ensemble. The HTyrAla$^{+}$ ions were nearly
completely destroyed and fragments of different mass have appeared.
For comparison, the red curve is the extraction spectrum of a crystal
acquired directly after the preparation phase (no R-IRMPD), showing
only the two ion species as expected (this spectrum is the same as
the blue curve in Fig.\ref{Flo:Hot vs. cold a} a)).

\begin{figure}
\begin{centering}
\includegraphics[width=15cm]{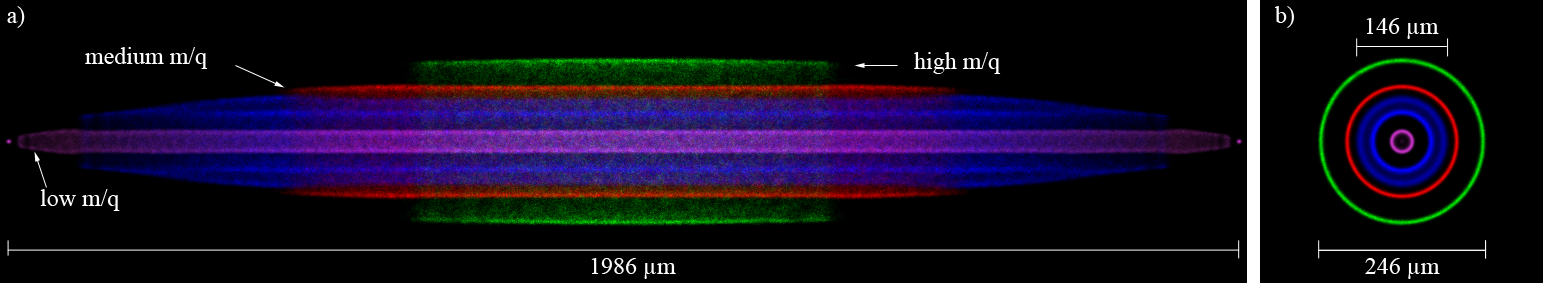}
\par\end{centering}

\caption{Mass separation in a multi-species Coulomb crystal. Simulated images
for the experimental trap parameters of a Coulomb crystal containing
440 molecular fragments (pink, mass 65\,amu), 1190 Ba$^{+}$ ions
(blue, 138\,amu), 460 BaO$^{+}$ ions (red, 154\,amu), and 230 HTyrAla$^{+}$
ions (green, 251\,amu). \textbf{a}) Side view. \textbf{b)} Cross
sectional view. Ions of lower mass arrange closer to the trap axis,
the heaviest ions arrange in a shell around the lighter ions.}

\label{Flo:Ion seperaton}
\end{figure}

The information contained in the ion extraction spectrum together
with the CCD images of the Barium ions' fluorescence before and after
extraction of the ions can be used to characterize the initial ion
ensemble (initial numbers of ions of different species and their temperatures)
and the final state (number of fragments and final temperature of
different species) \cite{Offenberg2008}. MD simulations are performed
for this purpose. Figure \ref{Flo:Ion seperaton}
 shows an example of the final state of an ensemble where R-IRMPD
has partially photodissociated the HTyrAla$^{+}$ ions. A comparison
with experimental CCD images reveals that the typical temperature
of HTyrAla$^{+}$ is less then 800\,mK. This value is the thermal
energy associated with the secular motion. In addition, all ions perform
micromotion. The micromotion energy of the HTyrAla$^{+}$ ions can
be calculated from the trap parameters and the distance from the axis,
and amounts to $k_{B,}(11,5$\,K) per ion \cite{Berkeland1998-Minimization}.

\section{IR laser source}

\begin{figure}
\begin{centering}
\includegraphics[width=12cm]{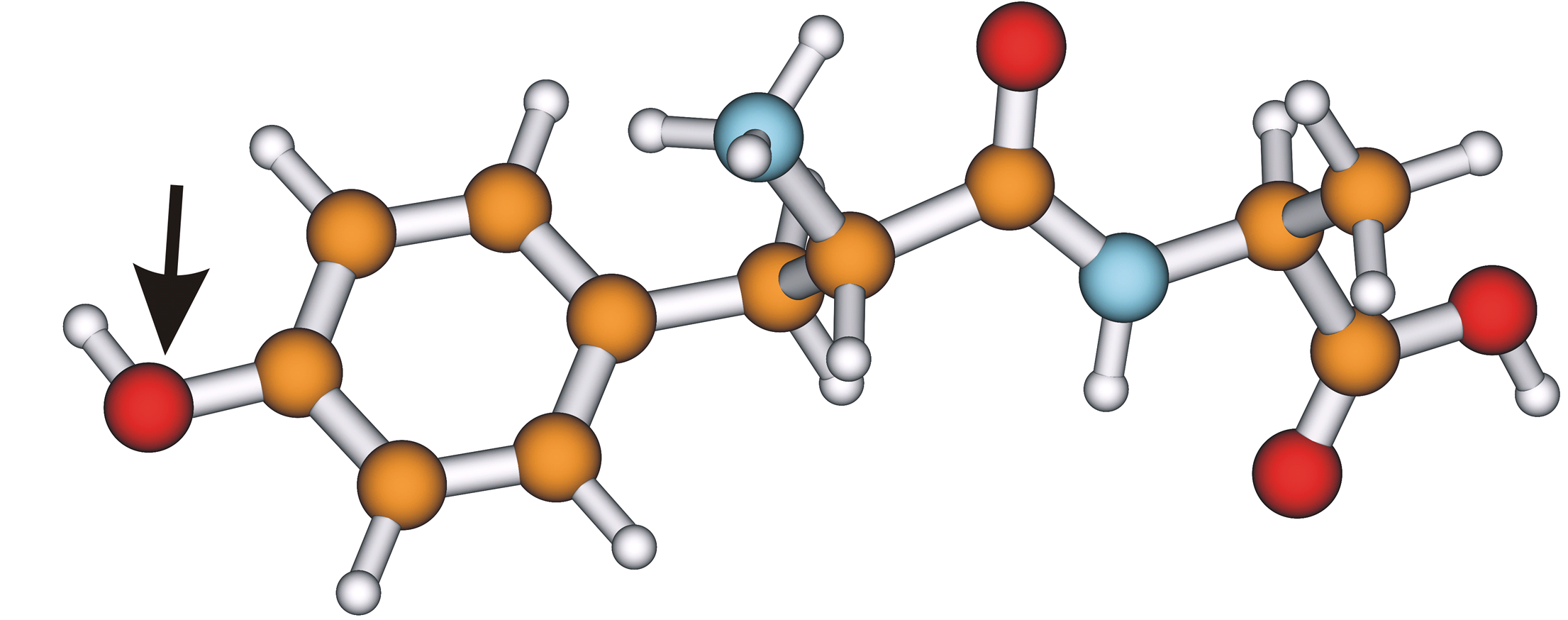}
\par\end{centering}

\caption{3D view of the structure of HTyrAla$^{+}$ (conformer A). Atoms: Carbon
(orange), oxygen (red), nitrogen (blue) and hydrogen (white). The
arrow indicates the phenol OH group in which the vibrational excitation
studied here takes place.}

\label{Flo:IR spectroscopy}
\end{figure}

HTyrAla$^{+}$molecular ions have a complex absorption spectrum, with
many lines corresponding to fundamental and overtone excitation of
local vibrational modes, complicated by the presence of a rotational
structure \cite{Stearns2007}. The single line studied here is the
fundamental vibration of the OH group bound to the phenol ring in
the HTyrAla$^{+}$ dipeptide (Figure \ref{Flo:IR spectroscopy}).
 We used the idler wave of a continuous-wave OPO for the spectroscopy,
since its high power and tunability provides experimental flexibility.
The device is home-built \cite{Vasilyev2010}. It uses a ring cavity
containing a periodically poled lithium niobate crystal, and is pumped
by an independent Nd:YAG master laser (1064\,nm), amplified by a
fiber amplifier, with a maximum output of 9.5\,W. Typically, 5\,W
are delivered via a fiber to pump the OPO ring cavity located next
to the UHV chamber (Figure \ref{Flo:overview trap}). This OPO is
tunable over a wide spectral range ($2500\ldots3000\,$nm) by changing
the temperature of the nonlinear crystal. The typical output spectral
linewidth is a few ten kHz fine, output power is up to 2.5\,W for
the idler wave and up to 3\,W for the signal wave (at the highest
pump power). In order to prevent mode-hops, the OPO is stabilized
using back refl{}ections from the gratings in the crystal. The output
wavelength of the idler radiation is determined by measuring the wavelength
of the signal radiation, which is sent to a high-resolution wavemeter.
The master laser is intrinsically frequency-stable on the time scale
of interest here, its wavelength being 1064.49\,nm. The idler wave
of the OPO is sent into the trap axially, by an uncoated CaF$_{2}$
beam sampler. Power is adjusted by a combination of a polarizer and
a rotated $\lambda/2$ wave retarding plate. The diameter of the idler
beam inside the trap was measured as 0.86\,mm.

\section{Calibration}

\begin{figure}
\begin{centering}
\includegraphics[width=9cm]{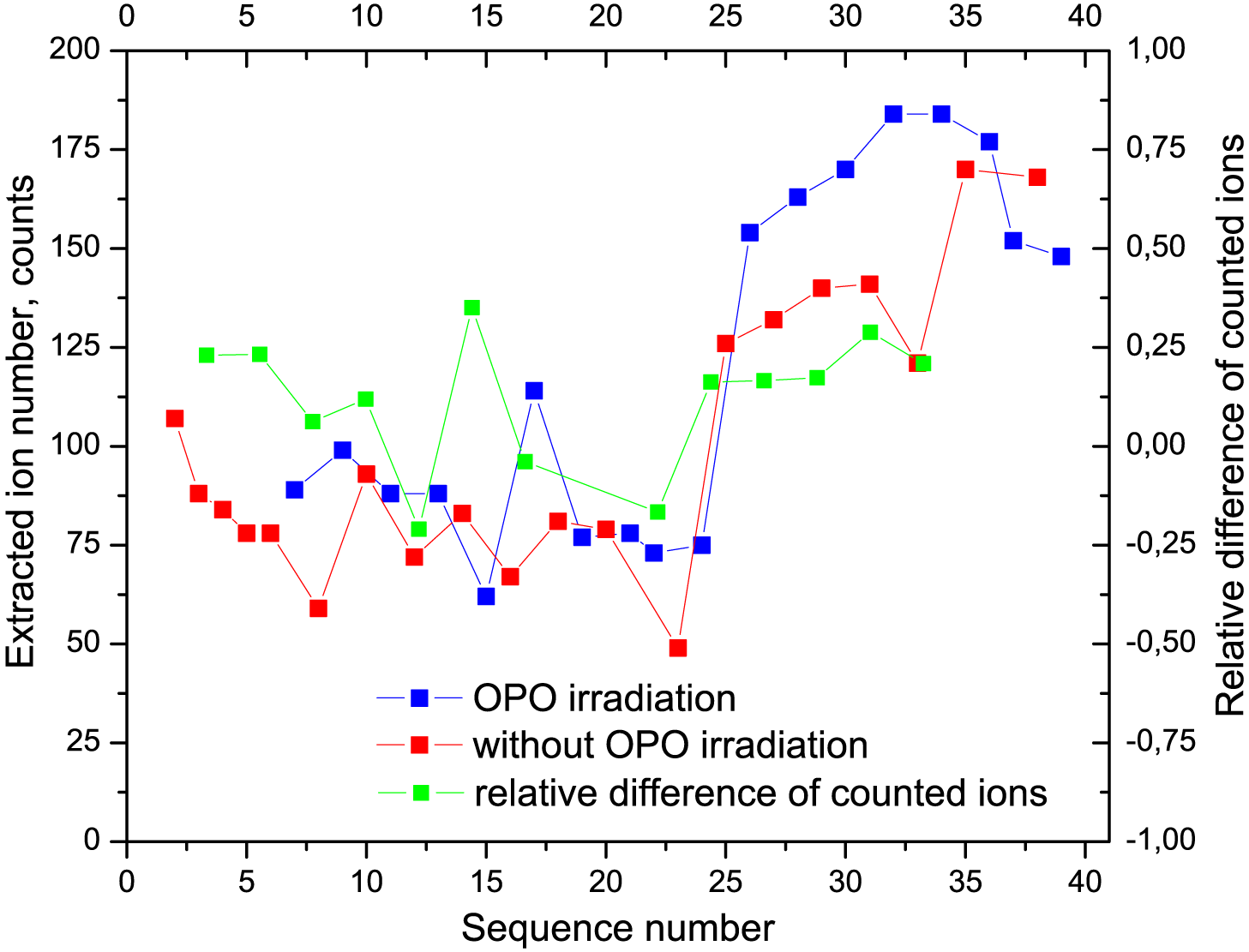}
\par\end{centering}

\caption{Ion flux check and ion number conservation. Red: counts of extracted
HTyrAla$^{+}$ without OPO irradiation. Blue: total counts of extracted
HTyrAla$^{+}$ and its fragments. Green: Relative difference of counted
ions (right ordinate). The comparison of the two cases shows that
the total count is equivalent to the number of loaded HTyrAla$^{+}$.}

\centering{}\label{Ion number conservation}
\end{figure}

The molecular ion flux from the ESI source is not stable, showing
both short-time fluctuation and longer-term trends, the overall flux
variation being on the order of 25\,\%. As a consequence, the number
of trapped HTyrAla$^{+}$ ions varied from one loading to the next.
In order to characterize the variations, the extraction spectra were
recorded during a series of repeated HTyrAla$^{+}$ loadings without
Ba$^{+}$. Interleaved procedures were used, alternating between OPO
irradiation on and off. Also, the OPO power and its idler wavelength
were changed within a wide range. Figure \ref{Ion number conservation}
 shows the measurements, taken over the course of 20\,min. Here,
the Barium ion crystal remained the same, laser cooling was always
on. The red points show the variation of the number of counts of extracted
HTyrAla$^{+}$ ions. For comparison, the blue points show the interleaved
measurements where after loading and buffer gas cooling, the HTyrAla$^{+}$
was partially photodissociated. Here, the number of counts of the
fragments and of the undissociated HTyrAla$^{+}$ in the mass spectrum
was added. The relative deviation (green) of this total number with
respect to the average number of HTyrAla$^{+}$ detected in the previous
and subsequent loadings is moderate, less than 25\,\%. Thus, this
total number of counts is used as a measure of the initial number
of HTyrAla$^{+}$ after loading, which is not measurable in a nondestructive
way, and also determines the effect of a spectroscopic excitation
in terms of the relative photodissociation efficiency of the initial
(parent) HTyrAla$^{+}$ ions. This procedure is quite useful, and
more practical than extracting the initial number from MD simulations,
which would be very time-consuming.

The typical ion extraction spectrum in Figure \ref{Flo:extraction b}
shows an advantage of having sympathetically cooled ions rather than
uncooled ions: this method allows a quantitative characterization
of the fragment ions. It is seen that there are at least three different
types of products of photodissociation, having mass/charge ratios
41, 65 and 90\,amu/e. Fragments with atomic mass close to $\frac{m}{q}=$138\,amu/e
are hard to distinguish from $^{138}$Ba$^{+}$. We assume that there
is no fragment with this mass which is proofed by experiments with
room-temperature ions. 

A series of R-IRMPD experiments was made for room-temperature and
sympathetically cooled ions, for different Coulomb crystal sizes and
numbers of HTyrAla$^{+}$ ions trapped, for different OPO wavelengths,
irradiation times and OPO powers. The extraction spectra were sorted
into groups corresponding to different power ranges, OPO frequencies
and irradiation times.

\section{Results and Discussion }

\begin{figure}
\begin{centering}
\includegraphics[width=8cm]{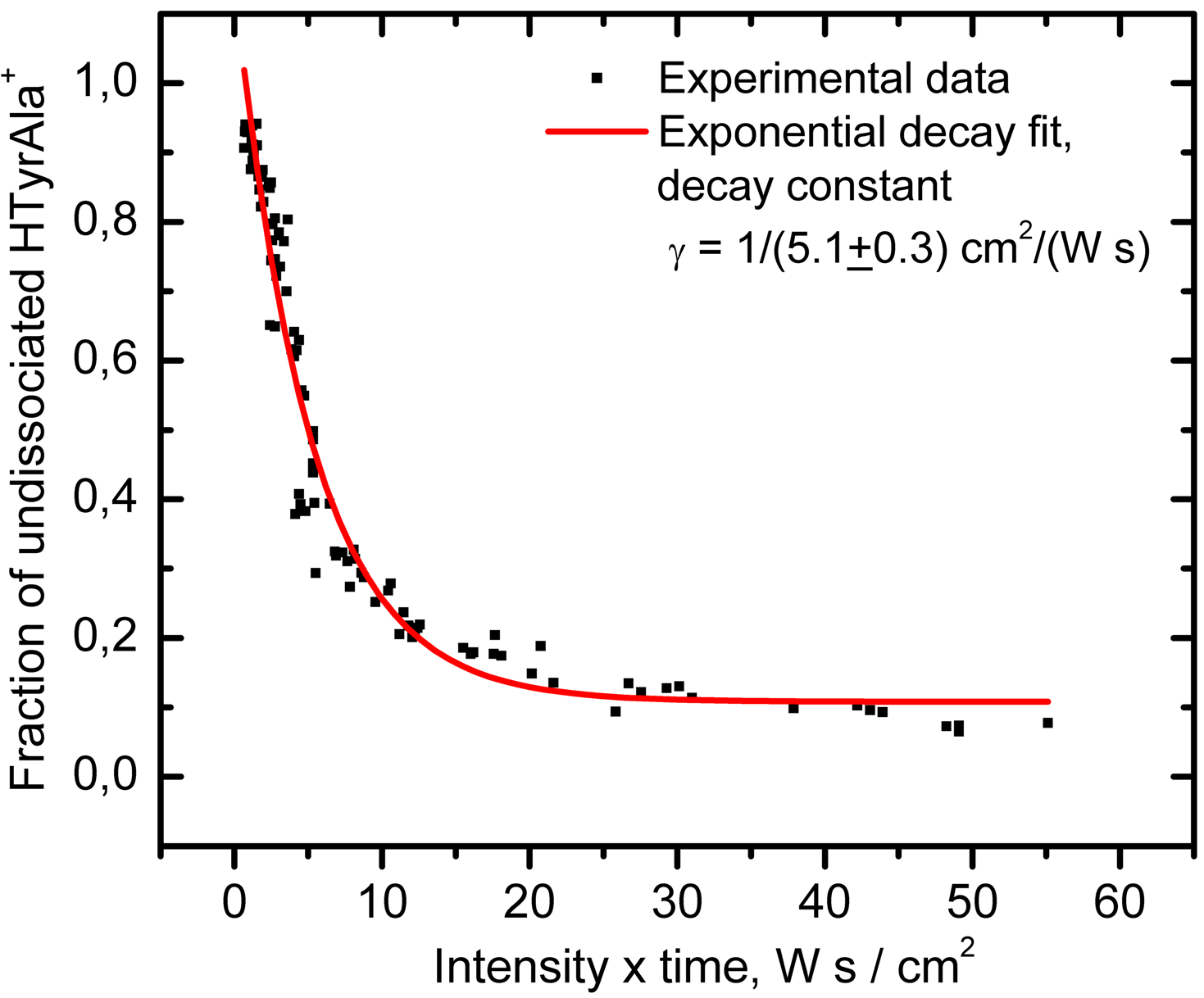}
\par\end{centering}

\caption{Fraction of undissociated uncooled HTyrAla$^{+}$ ions as a function
of irradiated energy. In this series of experiments, only HTyrAla$^{+}$
ions were trapped and cooled by 300 K He buffer gas, and irradiated
for different times and at different OPO power levels. As the measurements
were done on uncooled ions, significant broadening of the extraction
spectra occurs (see Fig. \ref{Flo:Hot vs. cold a}). This limits the
resolution of this method and gives rise to an additive noise. Therefore,
a fit function with an offset has been used to model the data.}

\label{Flo:complexvs.power}
\end{figure}

Figure \ref{Flo:complexvs.power} 
 shows the fraction of undissociated HTyrAla$^{+}$ ions as a function
of irradiated energy, given by the product of power and irradiation
time. In this series of experiments, only HTyrAla$^{+}$ ions were
trapped (in presence of He buffer gas), irradiated for different times
and powers (at the fixed bean diameter), while the OPO wavelength
was close to the central wavelength of the observed spectrum. The
fraction decreases exponentially with irradiated energy, indicating
that a linear process is occurring, where a one-photon process dominates.
Note that the dissociation rate is very small. At a typical power
of 10\,mW and the beam cross section used here, it is 0.34\,s$^{-1}$,
much smaller than spontaneous or IVR rate (see below). Taking into
account the beam cross section and correcting for the non-uniform
intensity distribution of the Gaussian OPO beam, we can deduce a (peak)
R-IRMPD dissociation cross section of at least $\sigma_{diss,min}=1.5\cdot10^{-20}$\,cm$^{2}$
under our conditions. As the buffer gas ion cloud may have had a larger
diameter than the laser beam, thus reducing the effective intensity
experienced by the ions, this value is a lower limit. On sympathetically
cooled ions, which are all located within the OPO wave, only a few
measurements of dissociation as a function of irradiation time were
performed at low intensity. The obtained decay constant is consistent
with that given in Fig.\ref{Flo:complexvs.power}, within the errors.
Thus, $\sigma_{diss}$ is similar to $\sigma_{diss,min}$.

\begin{figure}
\begin{centering}
\includegraphics[width=10cm]{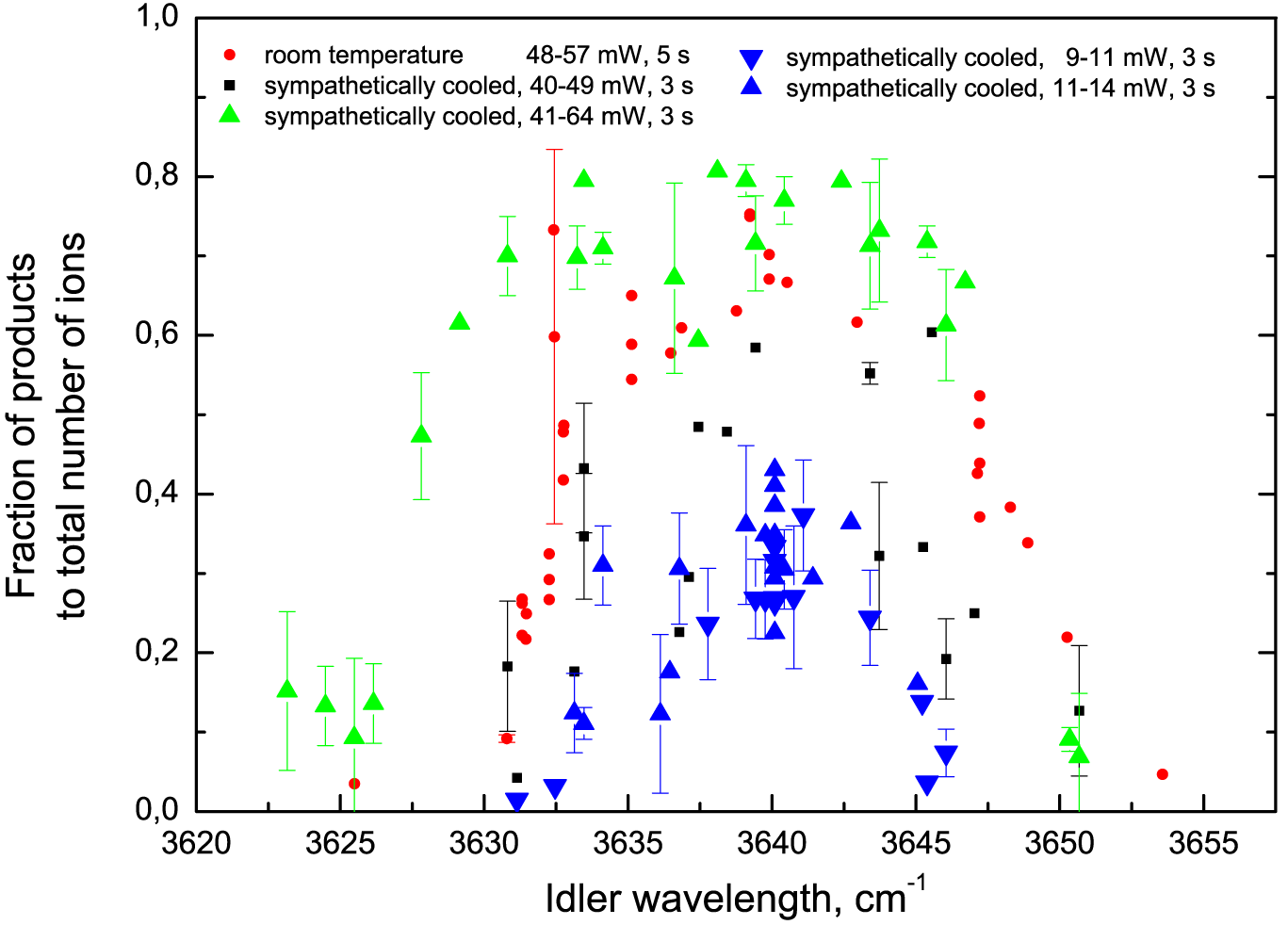}
\par\end{centering}

\caption{Spectra of room-temperature (buffer-gas cooled) and laser-cooled photodestructed
HTyrAla$^{+}$. The error of the room temperature (red) fraction of
products is significantly larger than for the sympathetically cooled
cases (green, black, blue).}

\label{Flo:Spektrum2}
\end{figure}

As phenol is that part of HTyrAla$^{+}$that carries the OH group
excited by R-IRMPD, we can compare $\sigma_{diss}$ with the peak
absorption cross section deduced from a measurement of the same mode
on phenol molecules, $\sigma_{PhOH,abs}=7.4\times10^{-19}$\,cm$^{2}$
\cite{Ishiuchi2006}. Here we have used the experimental oscillator
strength value $f=7.9\times10^{-6}$, and assumed a Lorentzian band
shape with a FWHM of 6\,cm$^{-1}$, obtained from a rotational band
simulation at 300\,K (see below). We can also consider the theoretical
absorption cross section for our particular vibrational transition,
obtained from the transition dipole moment of the mode as calculated
using Gaussian03 (see Table \ref{Table I}), and assuming the same
FWHM. For the four conformers the values are similar, $\sigma_{abs}=2\times10^{-18}\,$cm$^{2}$(peak
value), approx. three times the phenol value. Thus, $\sigma_{abs}$
is approximately a factor $10^{2}$ larger than the R-IRMPD cross
section $\sigma_{diss}$. A simple interpretation of this ratio is
that the molecule must absorb on the order of $10^{2}$ photons in
order to dissociate. Note that the corresponding total energy is approximately
one order of magnitude larger than the UV dissociation energy ($E_{diss}$
approx. 35\,000\,cm$^{-1}$ \cite{Stearns2007}). Two effects may
contribute to this factor. First, the excited molecule may fluoresce,
either from the excited state, or from states that it evolves into
as a consequence of IVR. In a very simple model, the ratio of decay
rate $\Gamma$ from the excited state via single or multiple photon
emission and the total decay rate $A+\Gamma$, where $A$ is the fluorescence
rate, determines the actual energy deposition rate, $dE_{int}/dt=\sigma_{abs}I\,\Gamma/(\Gamma+A)$,
where $I$ is the intensity (stimulated emission is neglected). The
ratio may be distinctively smaller than unity and contribute to a
reduction of $\sigma_{diss}=(\Gamma/(A+\Gamma))(h\nu/E_{diss})\sigma_{abs}$.
The fluorescence rate $A$ for the excited vibrational state is calculated
to be approximately 95\,s$^{-1}$, using the transition dipole matrix
elements given in Tab.\ref{Table I} below. (The calculated stimulated
emission rate at the OPO intensity shown in Fig.\ref{Flo:Spektrum2}
(blue curves) is approximately 50\,s$^{-1}$, smaller than $A$).
Second, the molecules, internally heated by energy deposition, will
also {}``cool'' by emission of black-body radiation \cite{Dunbar1992}.
Since the time scale over which dissociation is obtained under our
experimental conditions is long (of order 1 s) this could be a process
of relevance. Such a cooling process would also increase the number
of IR photons necessary to be absorbed until sufficient energy is
accumulated that leads to dissociation. A detailed explanation of
the ratio $\sigma_{diss}/\sigma_{abs}$ is beyond the scope of this
work, requiring extensive additional measurements.

The experimentally observed photodissociation spectra are shown in
Figure \ref{Flo:Spektrum2}. 
 The fundamental transition of the phenol-OH mode has the central
wavelength of approx. 3640\,cm$^{-1}$, within a few wavenumbers
both for buffer-gas and sympathetically cooled ions. The measurements
shown in blue are taken at the lowest intensities, and the maximum
dissociation fraction is below 0.4. This implies that saturation effects
are not important in this case. The full width of this line (for sympathetically
cooled ions) is approx. 10\,cm$^{-1}$ and will be discussed in the
following.

The measurements taken with buffer gas cooled molecules show a larger
linewidth. This is likely due to both saturation (a larger OPO intensity
was used) and to an increase of internal energy (rotational temperature)
caused by partial transformation of micromotion energy, which is much
larger for buffer-gas cooled ions than for sympathetically cooled
ones, to internal energy during collisions with He atoms.

The observed transition linewidths contain information about the internal
rotational temperature. However, this information cannot be extracted
straightforwardly, as several unknowns contribute to it. First, the
preparation step of our cold molecules is not sensitive to conformer
\cite{Stearns2007} having overall energies close or below the level
$k_{B}\times$(room temperature), so any ensemble of ions loaded into
the trap is expected to contain various conformations and these will
contribute to our spectroscopic signal. Different conformations have
different rotational constants and vibrational transition frequencies
(see 
\begin{table}
\begin{tabular}{|c|c|c|c|c|c|c|c|c|}
\hline 
\begin{sideways}
r
\end{sideways} & \multicolumn{6}{c|}{} & \multicolumn{2}{c|}{}\tabularnewline
\begin{sideways}
e
\end{sideways} & \multicolumn{6}{c|}{Theoretical} & \multicolumn{2}{c|}{Experimental}\tabularnewline
\begin{sideways}
m
\end{sideways} & \multicolumn{6}{c|}{} & \multicolumn{2}{c|}{}\tabularnewline
\cline{2-9} 
\begin{sideways}
r
\end{sideways} &  &  &  & \multicolumn{3}{c|}{} &  & \tabularnewline
\begin{sideways}
o
\end{sideways} & Energy rel. & Vibr.  & Transition dipole & \multicolumn{3}{c|}{Rotational constants } & Vibr. & Line-\tabularnewline
\begin{sideways}
f
\end{sideways} & to ground  & freq. of & moment & \multicolumn{3}{c|}{{[}GHz{]}} & freq. & width\tabularnewline
\begin{sideways}
n
\end{sideways} & {[}kJ/mol= & PhOH mode & $(d_{x},d_{y},d_{z})$ & \multicolumn{3}{c|}{} & {[}cm$^{-1}${]} & {[}cm$^{-1}${]}\tabularnewline
\begin{sideways}
o
\end{sideways} & $k_{B}$(121K){]} & {[}cm$^{-1}${]} &  {[}$0.0104\,$D{]} & \multicolumn{3}{c|}{} &  & \tabularnewline
\cline{5-7} 
\begin{sideways}
C
\end{sideways} &  &  &  & A & B & C &  & \tabularnewline
\hline 
\hline 
A & 0 & 3641.2 & (9.96, 2.42, 3.59) & 1.0902 & 0.1442  & 0.1347  & 3641.5  & 2\tabularnewline
\hline 
B & 4 & 3642.2 & (3.15, 8.60, 5.06)  & 0.6949  & 0.2018  & 0.1792  & 3642  & 2\tabularnewline
\hline 
C & 0.1 & 3641.8 & (-8.55, 4.35, 5.12)  & 1.0912  & 0.1444  & 0.1346  & 3642  & 2\tabularnewline
\hline 
D & 4.1 & 3642.8 & (10.6, 1.16, 1.56)  & 0.6911  & 0.2022  & 0.1795  & 3642.5  & 2\tabularnewline
\hline 
\end{tabular}

\caption{Low-energy conformers of HTyrAla$^{+}$: theoretical and experimental
properties from \cite{Stearns2007}. The transition dipole moment
components are given in the principal axis system. The experimental
spectrum of \cite{Stearns2007} was taken in steps of 0.5\,cm$^{-1}$,
thus the differences between the conformers' frequencies are only
known with this resolution. The linewidth values are approximate.
The observed vibrational frequency has a systematic uncertainty of
2\,cm$^{-1}$. The theoretical vibrational frequencies were obtained
from quantum chemical computations and corrected by a factor 0.954
in order to take into account anharmonicities of hydride frequencies.
This factor has previously given a good fit for protonated amino acids.}

\label{Table I}
\end{table}
 Table \ref{Table I}). This introduces a broadening and shape change
of the vibrational line. Second, in R-IRMPD, the molecule absorbs
many photons before dissociating. The increasing internal energy may
lead to an increasing distortion of the molecular structure. As this
occurs, the vibrational frequency of the local mode may shift, yielding
another cause of broadening. Third, the molecule may be heated rotationally.
Fourth, the IVR process leads to line broadening of each ro-vibrational
transition line in the band, washing out the fine structure of the
rotational band. 

The measurements of \cite{Stearns2007} on HTyrAla$^{+}$ provide
important information on these issues. In their work, vibrationally
cold ions were obtained by collisions with cold He buffer gas atoms
in a cryogenic trap. The vibrational temperature was measured to be
10-11\,K and the roational temperature was presumed to be in equilibrium.
Their measurements include information on the relative shift of the
vibrational lines for four conformers and also on the rotational band
linewidths (see \ref{Table I}). It should be emphasized that their
spectroscopic method was not R-IRMPD but vibrationally depleted electronic
(UV) spectroscopy, in which only a single IR photon is absorbed before
the following dissociation, therefore a distortion-induced lineshift
cannot contribute to their observed linewidths. As their work was
focused on determining a wide IR spectrum rather than precisely measuring
one particular line, a frequency step size of 0.5\,cm$^{-1}$ was
suffcient. Their measurements indicate that the frequency shift between
conformers is not more than about 1\,cm$^{-1}$ roughly consistent
with the theoretical relative shifts. The FWHM linewidth of individual
conformer species is about 1.7\,cm$^{-1}$, after subtracting the
contribution of the spectral linewidth of the pulsed OPO (linewidth
ca. 1\,cm$^{-1}$). Importantly, this linewidth includes the effect
of IVR. Our 10\,cm$^{-1}$ linewidth (in the sympathetically cooled
case) is substantially larger than both the {}``cryogenic'' linewidth
and the upper limit for the conformer shift. Therefore, in our sympathetically
cooled case conformation variations and IVR broadening cannot be dominant
effects. Instead, either our rotational temperature is significantly
higher than 10\,K or there is a substantial internal-energy-induced
line shift, or both. 

\begin{figure}
\begin{centering}
\includegraphics[width=8cm]{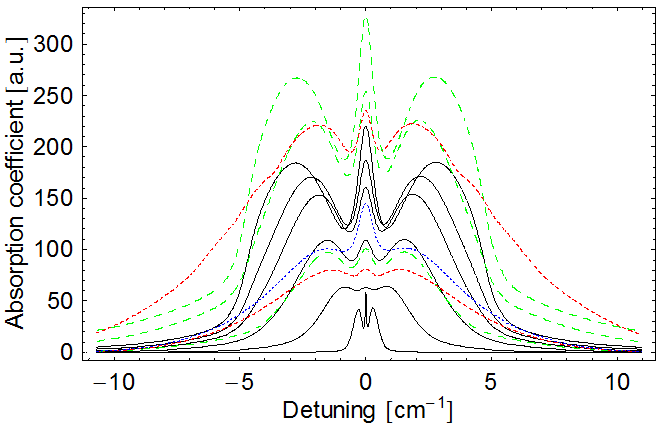}
\par\end{centering}

\caption{Simulated rotational bands of the phenol OH vibration of HTyrAla$^{+}$.
Full lines: conformer A. The temperatures are, increasing from the
smallest line, 10\,K, 100\,K, 300\,K, 450\,K, 600\,K, 1000\,K.
The red/medium-long-dashed (green/long-dashed, blue/short-dashed)
lines are for the B (C, D) conformer. The FWHMs are as follows. Conformer
A: 1\,cm$^{-1}$(10\,K), 3.9\,cm$^{-1}$(100\,K), 6.2\,cm$^{-1}$(300\,K),
7.3\,cm$^{-1}$(450\,K), 8.1\,cm$^{-1}$(600\,K), 9.2\,cm$^{-1}$(1000\,K);
conformer B: 8.5\,cm$^{-1}$(300\,K), 11\,cm$^{-1}$(600\,K);
conformer C: 6.5\,cm$^{-1}$(300\,K), 8.4\,cm$^{-1}$(600\,K),
9.2\,cm$^{-1}$(1000\,K); conformer D: 6.5\,cm$^{-1}$(300\,K).
The vertical scale is arbitrary for all curves, also for different
temperatures of a single conformer.}

\label{Flo:PGOPHER}
\end{figure}

In order to discuss the influence of temperature on the band linewidth,
simulations of the rotational band for different rotational temperatures
were made. As input parameters the rotational constants as well as
the vibrational transition dipole moments in the molecular frame,
as computed by \cite{Stearns2007} and reported in Table \ref{Table I}
were used. The rotational constants of the vibrational ground state
are also used for the excited state, due to lack of any further information.
The rotational band structure was computed using the program \emph{PGOPHER}
\cite{PGOPHER}, including transitions between rotational levels with
quantum number up to $J=510$. This high value was required in order
to obtain a sufficiently complete band shape even at the highest considered
temperature, $T=1000\,$K, at which a very large number of rotational
levels in the ground state are thermally populated. The individual
ro-vibrational transitions were broadened by a Lorentzian line shape
function of 0.9\,cm$^{-1}$ full width at half maximum, in order
to obtain a smooth band. A value smaller than the contributions from
IVR and from the presence of more than one conformer(s) was chosen,
in order not to smooth too much any thermal features. Figure \ref{Flo:PGOPHER}
 shows the simulated spectra for different rotational temperatures.
The resulting line shapes are not Gaussian. The simulated spectrum
for 10\,K (where a reduced broadening of 0.067\,cm$^{-1}$ has been
used in order to show the structure of the band more clearly) has
a width of 1\,cm$^{-1}$, which can be compared with the measurements
of \cite{Stearns2007}, (approx. 1.7\,cm$^{-1}$). The simulated
width is smaller, but an additional contribution coming from IVR broadening
must be considered as well. Yamada et al. \cite{Yuji2004,Yamada2007}
have measured the lifetime of the upper vibrational state of the OH
vibration in phenol as 14\,ps, which corresponds to a linewidth of
0.4\,cm$^{-1}$. It is conceivable that the lifetime is shorter in
HTyrAla, due to its larger density of states as compared to that of
phenol, which may provide more doorway states. The IVR rate strongly
depends on such states, as has been shown by Yamada et al., who find,
e.g. a six times smaller rate when the phenol ring is deuterated (phenol-$d_{5}$),
which significantly changes the vibrational frequencies of the molecule,
and a three times higher rate in case of the OH vibration. Such effects
may explain the slightly larger linewidth value observed by Stearns
et al.

For rotational temperatures from room temperature on, the simulated
linewidths increase only slowly with temperature, e.g. at $T=(100,\,300,\,450,\,600,\,1000)\,$K
the simulated FWHMs are approximately $(3.9,\,6.2,\,7.3,\,8.1,\,9.2)\,$cm$^{-1}$
for conformer A. Thus, in principle the FWHM can be correlated with
rotational temperature, but obtaining a high temperature resolution
would require an accurate measurement of the bandshape. This was not
possible in our work, due to variations in the ion flux and the limited
signal-to-noise ratio achieved with of the low trapped ion number.
For comparison, conformers B, C, D were also simulated, which have
linewidths $(8.5,6.5,6.5)\,$cm$^{-1}$ at 300\,K, respectively.
Conformer B has a significantly larger FWHM at room temperature and
above than conformer A and C. This variability further makes an accurate
determination of rotational temperature difficult in the case of a
mixture of conformers. The observed FWHM of 10\,cm$^{-1}$ corresponds
to an effective rotational temperature in the range from above 600\,K
to above but near 1000\,K, significantly above the 300\,K that the
molecules are expected to have after collisions with 300\,K Helium
buffer gas.

\section{Conclusion}

Vibrational spectroscopy of a local mode of a sympathetically cooled
polyatomic molecular ion was demonstrated, for the first time to our
knowledge. The photodissociation of polyatomic HTyrAla$^{+}$ ions
was induced by irradiation with continuous-wave infrared light at
2.74\,\textmu{}m via the R-IRMPD process. For comparison, experiments
were also performed on uncooled ions. The kinetic energy of the ions
was significantly different in these two cases. After loading of the
ions into the trap, room-temperature He gas was introduced as a first
translational cooling step. In the case of subsequent sympathetic
cooling, the rotational temperature may be assumed to have reached
a value on the order of 300 K, since interaction with the 300 K black-body
radiation is dominant. Without sympathetic cooling, the ion micromotion
energy may have been partially converted to internal energy, leading
to an increased rotational temperature. 

For the lower IR intensities used, a photodissociation rate was observed
that is linearly dependent on intensity, and allowed determination
of the appproximate R-IRMPD cross section. It is significantly (approx.
two orders) lower than the vibrational absorption cross section, indicating
that a significantly larger number of IR photons must be absorbed
before dissociation can take place, than that corresponding to a typical
bond energy. The R-IRMPD spectral lines exhibited modest signal-to-noise
ratios, limited by the fact that the individual dissociation experiments
are performed with small number of ions and this number fluctuates
from loading to loading. By comparison with measurements on HTyrAla$^{+}$
cooled by cold He buffer gas by Stearns and coworkers, we deduce that
the observed linewidth of the vibrational transition (10 cm$^{-1}$)
is not dominated by IVR from the upper vibrational state, nor by the
small frequency shifts among conformers that are present as a mixture
in our molecular samples. Instead, it is due to a combination of rotational
temperature imposed by the initial preparation, the interaction with
black-body radiation of the environment, a possible increase of rotational
temperature in the course of absorbing the approx. 80 IR photons,
and a shift of the vibrational frequency with increasing internal
energy in the molecule. The latter would be caused by cross-coupling
anharmonicities of the molecule's modes. Phenomenologically, the effective
temperature to be assigned to the line is significantly above room
temperature, approximately 1000\,K. Further studies are required
to disentangle the mentioned effects. 

Future investigations could be aimed at (i) an accurate determination
of the IVR-induced broadening, by performing measurments as in \cite{Stearns2007},
however, using an injection-seeded pulsed OPO with reduced linewidth
or a continuous-wave OPO, (ii) determine more precisely the influence
of internal-energy-induced broadening, by performing R-IRMPD on rotationally
cold HTryAla$^{+}$, in a cryogenic trap apparatus (as used by Stearns
et al.), and measuring with high signal-to-noise ratio the rotational
band shape. As for (i), the measurements should be performed on samples
containing a single conformer, which can be obtained by selectively
UV-dissociating other conformers. Moreover, a combination IR-multiphoton
absorption and IR-depletion spectroscopy could be applied: the vibrational
lineshape is measured by IR-depletion spectroscopy using an injection-seeded
pulsed OPO with reduced linewidth, but preceding the interrogation
by multi-photon IR excitation for a fixed but variable time, under
conditions that the molecules mostly remain undissociated. In this
way, one could determine the line shift as a function of internal
energy. One could also measure the vibrational temperature as a function
of internal energy, using hot band intensities as indicator. 

In such measurements, the cryogenic buffer gas should first be removed
from the trap in order not to continue rotational cooling. Under such
conditions, the molecules would still be rotationally and vibrationally
cooled by IR emission, but with reduced rate. A determination of this
cooling rate is itself an interesting aim.

\section{Acknowledgments}

The authors are grateful to T.~Schneider for helpful discussions
and to B.~Roth for his contributions in the initial phase of this
project. We especially acknowledge T. Rizzo, J. Stearns, and T. Wassermann
for important suggestions, discussions and communication of unpublished
results. This work was funded by the DFG under grant SCHI 431/12-1.

\end{document}